\documentclass[12pt,a4paper]{article}
\usepackage[english]{babel}\usepackage{hyperref,epsf}\newcommand{\N}{\noindent}
\begin{document}
\title{Exciton condensation in semiconductor quantum wells\\in
nonuniform electric field}
\author{A.A.~Chernyuk~$^1$, V.S.~Kopp~$^2$, V.I.~Sugakov~$^1$ \\
$^{1}$\textit{Institute for Nuclear Research, Nat. Acad. of Sci. of Ukraine,} \\
\textit{\normalsize pr.~Nauky, 47, Kyiv-03680, Ukraine} \\
$^{2}$ \textit{Taras Shevchenko Kyiv National University,}\\
\textit{\normalsize pr.~Glushkova, 2, Kyiv-03127, Ukraine}\\
(e-mail: {\it sugakov@kinr.kiev.ua})}

\maketitle

\begin{abstract}{\small The structure appearance in exciton density
distribution in semiconductor double quantum well with transverse
electric field applied is studied, in the case when the metal
electrode contains a round window. It is suggested that there is
exciton condensed phase, free energy of which can be described by
phenomenological Landau model. For the exciton density determination
the traditional theory of phase transitions was used generated for
the case of the finite exciton lifetime, a presence of the pumping
and nonhomogenity of the system. It is shown that at high exciton
density the different types of structures appear: periodic
distribution of exciton condensed phase islands or condensed phase
rings. The behavior of the structures depending on the pumping, the
window size and temperature is analyzed. The obtained results are
agreed with experimental data.}\end{abstract}


\section{Introduction}

Interexciton interaction in semiconductor double quantum wells is
intensively studied, particularly in connection with a problem of
revealing Bose-Einstein condensation of excitons \cite{MoscSnoke}.
Important results were obtained under investigation of so-called
``indirect excitons'' in double quantum wells. In the case of
applying the electric field, recombination of indirect exciton is
hampered and the lifetime becomes long, therefore great
concentration of excitons can be created. In semiconductor alloys
based on GaAs/AlGaAs double quantum wells, a narrow luminescence
line was revealed at low temperatures, which had some specific
features: the line appears at pumping, larger than some threshold
value, which depends on temperature; the emission intensity of the
line depends super-linearly on pumping etc. \cite{LarTim,Tim2}. In
exciton luminescence spectrum in double quantum well the
luminescence from the laser spot region was revealed to be
surrounded by a concentric bright ring of emission, separated from
the central spot by a dark region \cite{ButovNat,SnokeNat}; the
distance between the ring and the laser spot enlarges with pumping
growth, and at low temperatures the ring was observed to be broken
down into periodically sited fragments \cite{ButovNat}.

There are different interpretations of the structure formation. A
series of explanations of the phenomena was connected with Bose
statistics of excitons. According to \cite{ButovPRL05}, the
mechanism, based on stimulated Bose character of exciton statistics
in relaxation process, gives rise to instability in the exciton
subsystem. In papers \cite{Liu,Parask} it was shown the possibility
of appearance of periodic structures in solutions of nonlinear
quantum equations like nonlinear Schrodinger equation. But the
investigation of observed pecularities, changes in the structure
depending on temperature, pumping and the parameters of exciton
system in \cite{Liu,Parask} was not considered.

Another alternative approach for the explanation of the
peculiarities in the system with great exciton density was developed
in [9--13]. The explanation is not connected with Bose statistics of
excitons. In these papers a traditional phase transition theory,
generalized for the case of unstable particles and a presence of the
pumping, is applied. The appearance and the properties of the
forming structures are related with: 1) the presence of the
attraction between excitons, which leads to condensed phase
formation (the possibility of condensed phase formation of indirect
excitons was shown in \cite{LozBerm}), and 2) non-equilibrium of the
system, caused by finite exciton lifetime.

As it was shown in \cite{Sug86}, the system of attracting excitons
is unstable with respect to density super-lattice formation. The
properties of exciton condensation in 2D systems in a statistical
approach, generalized for the case particles with finite lifetime,
were studied in \cite{Sug4}. Ring fragmentation outside of the laser
irradiation region in quantum wells observed in \cite{ButovNat} was
explained in a statistical approach \cite{Sug3} and in spinodal
decay model, generalized for the system of unstable particles
\cite{Ch,ChP}. The emerging structure in case of interaction of two
laser spots, was simulated in \cite{ChP}. The method employed in
\cite{Ch,ChP}, applied for the investigation of a nonuniform system,
allowed to explain the transition between fragmented luminescence
ring and a continuous one with temperature or other parameters
changing, the formation of localized spots in the emission, caused
by defects in the structure and etc. So, the appearance of a
structure in high-density exciton systems was explained in the
papers [9--13]with the processes of self-organization in
non-equilibrium systems caused by both the finite value of the
exciton lifetime and the presence of pumping. Such explanations do
not require the Bose-Einstain condensation of excitons.

Recently the interesting effects were observed in the papers
[16--18]. The excitation and emission of indirect excitons was
performed through ``a window'' in metallic gate with the diameter of
order of several micrometers. In the case of excitation intensity
growth, a regular ring pattern of equidistant bright spots along the
perimeter of the window was observed. At higher pumping or
temperatures the structure washed out and transforms to the emission
from a ring. With expanding window size, the structure became
complicated. These experiments are explained in this paper,
continuing the viewpoint of the papers [9--13].

\section{Model of the system}

Let us deal with qualitative picture of structure appearance of
exciton condensed phase in quantum well in the case of a window in
the electrode. Increasing the electric field, the position of the
indirect exciton level shifts towards low energies. Under the window
the electric field is smaller, than in the region remoted from the
window, therefore, a hump for exciton potential energy arises in the
well. Excitons, created by the light through the window, roll down
in nonuniform field from the middle of the well towards the region
under the window. Due to a finite lifetime they can not move far
away from the region under the window. As a result, in the well
under the perimeter of the window the maximum of the exciton density
appears. Thus, at the pumping growth the condensed phase forms in
the region of maximum exciton density, i.e., on the ring along the
perimeter of window, {\it like it was} observed in experiments
[16--18].

Let us find the exciton potential energy, caused by the window
presence in metallic electrode above the quantum well plane
(Fig.~1).

\N\epsfxsize=0.8\columnwidth\centerline{\epsffile{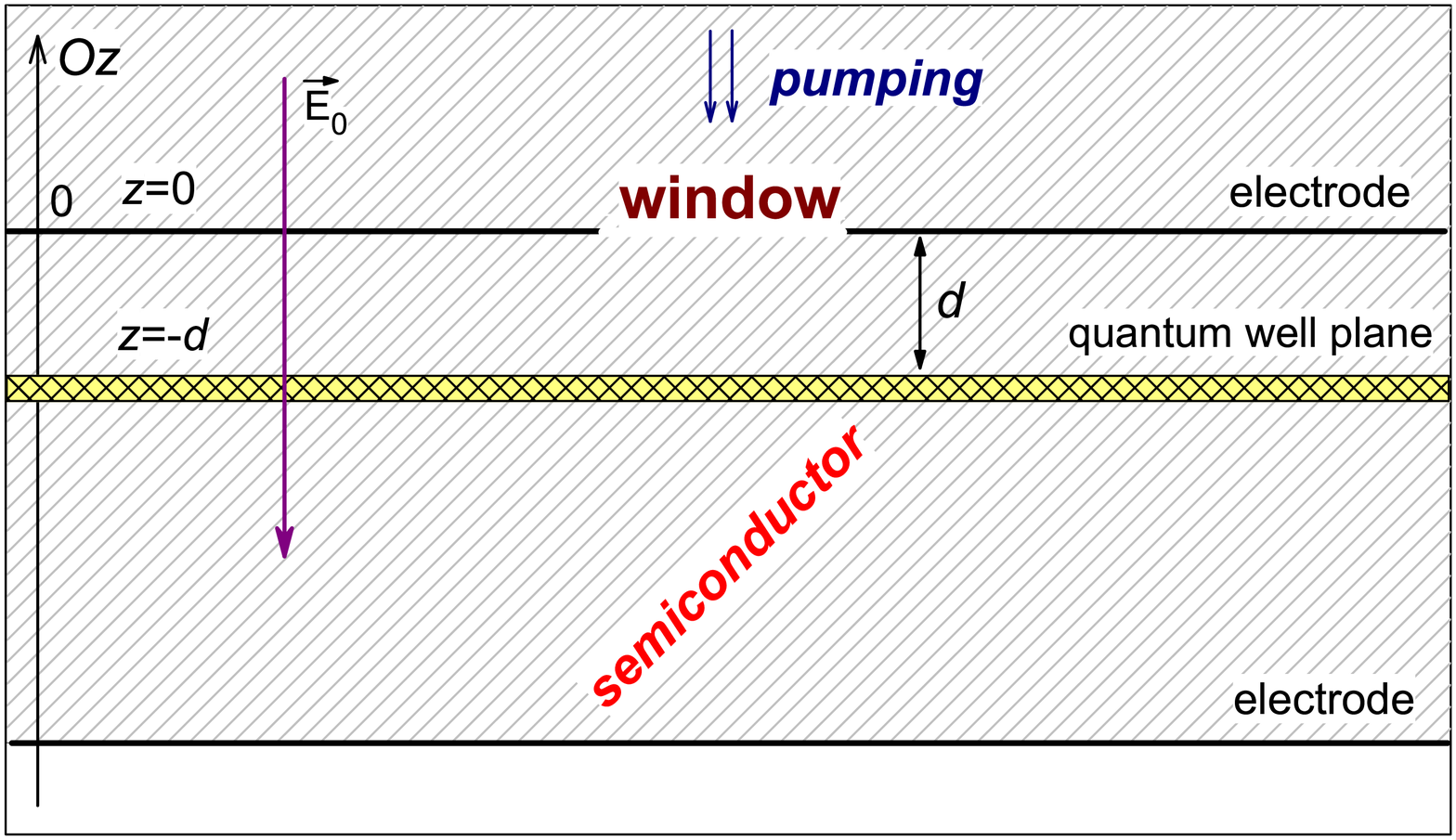}}

\centerline{\small Fig.~1. Scheme of the system.}

\N In external electric field an exciton gets additional energy
$V=-p_{\rm z}E_{\rm z}$, where $p_{\rm z}$ is the dipole momentum of
an exciton (the axis $Oz$ is perpendicular to the quantum well
plane). In experiments [16--18] semiconductor with double quantum
well is covered by a metallic mask, in which a round window is
located. The uniform electric field of the intensity $E_0$ is
applied perpendicular to the quantum well plane. Due to the presence
of the window in the mask, the electric field is distorted at
immediate proximity of the window. Let us calculate nonuniform
addition to external uniform electric filed into the exciton
potential energy. Let the upper electrode with the coordinate $z=0$
has a round window of the radius $r_0$. For the determination of the
field we have to solve Laplace equation for the potential with the
following boundary conditions: the potential of the electrostatic
filed is constant at both electrodes; the difference of the
potentials of electrodes ought to be equal to the value, at which
the field between electrodes is equal to $E_0$ far away from the
window. For that we shall use the solution of the task given in
\cite{LL} about the field, created by the ground plane with a
window, located in external electric field of the intensity $E_0$.
As a matter of fact, this solution does not satisfy the condition of
the constant potential on the lower electrode. But a change of the
potential introduced by the window falls down with moving off from
the window as a dipole potential \cite{LL}, and thus, it is small in
the region of the lower electrode under conditions that $r_0\ll L$,
where $L$ is the distance between electrodes. Let us consider that
$L\gg r_0$, $z$, i.e., the plane of the well is located
significantly nearer to the upper electrode, than to lower one.
Besides this, for using the solution of the task \cite{LL}, in which
the environment for both sides of the window is identical, we
consider, that the upper electrode (the electrode with the window)
is inner and is located inside semiconductor environment. So, the
considered system slightly differs from the investigated in the
experiments [16--18], but it gives the same qualitative behavior of
the potential at the change of the both the window radius and the
location of the well relatively the electrode.

\N Thus, the additional potential, created by the window is

\begin{equation}\label{phi}\varphi=\frac{E_0z}{\pi}\left(arctg
\frac{r_0}{\sqrt\xi}-\frac{r_0}{\sqrt\xi}\right),\end{equation}

\N where $\xi\equiv\frac{1}{2}\left[\rho^2+z^2-r_0^2+
\sqrt{\left(\rho^2+z^2-r_0^2\right)^2 + 4z^2r_0^2}\right]$ is
flattened spheroidal coordinate, $\rho$ is the radial coordinate in
the quantum well plane. At great fields, when the electrons and
holes are distributed to different wells and dipole momentum does
not depend on $E_0$, the exciton potential energy
$V\left(\rho,z\right)=p_{\rm z}\cdot
\partial \varphi/{\partial z}$. We shall characterize the value of
the potential energy by ``pulling force'' $\lambda=p_{\rm z}
E_0/\left(k T_{\rm c}\right)$. A radial profile of the exciton
energy $V\left(\rho\right)$ at different values of $z/r_0$ is
presented in Fig.~2.

\N\epsfxsize=0.8\columnwidth\centerline{\epsffile{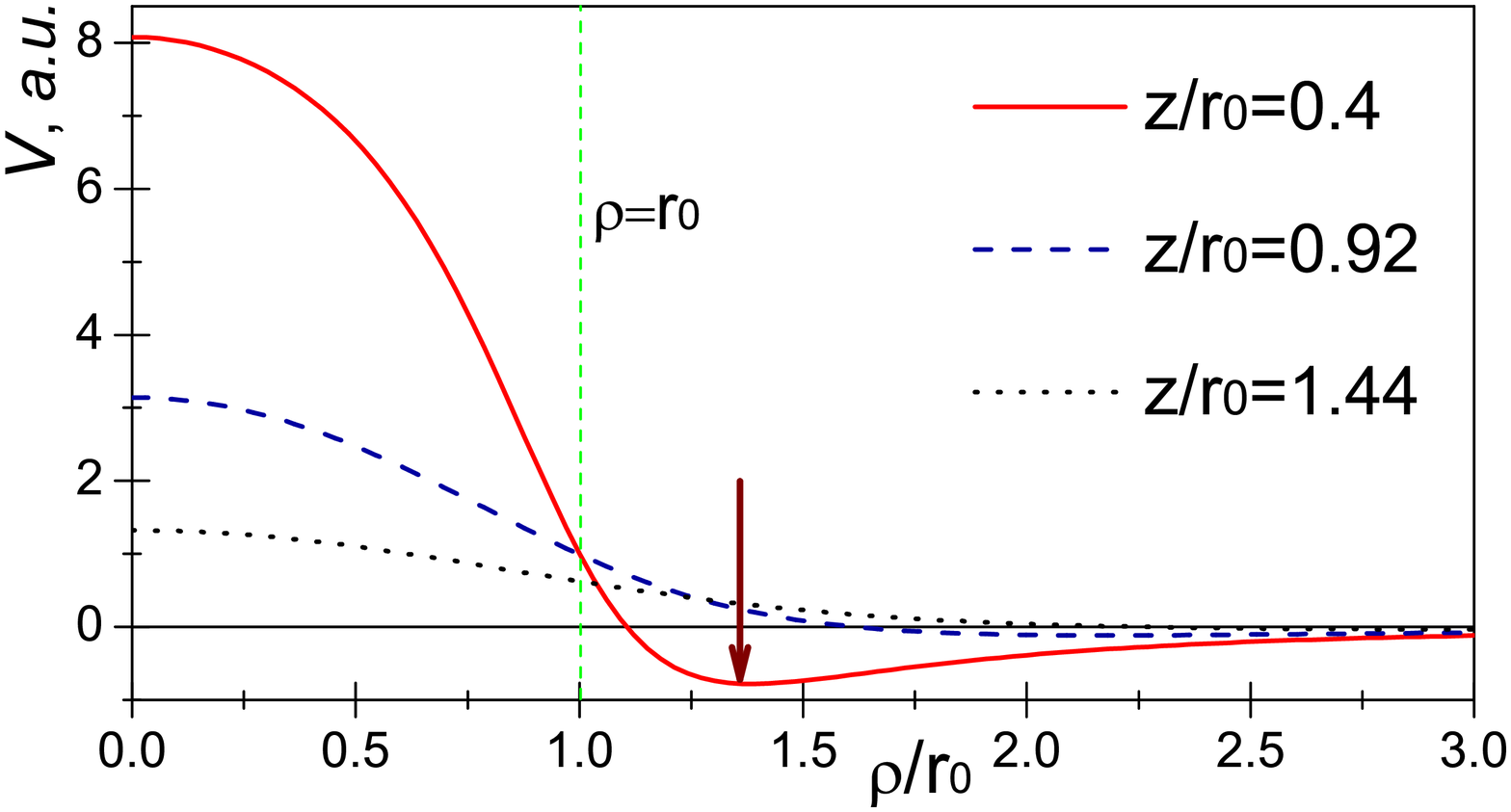}}

\N{\small Fig.~2. A radial profile of the exciton potential energy
$V$ in nonuniform electric field as a function of the ratio
$\rho/r_0$ at different values of the parameter $z/r_0$. The pulling
force $\lambda=30$ is for all curves}

\N Coordinate $z$ characterizes the distance from the quantum well
to the plane of the electrode with the window. With decreasing this
distance, the potential hump for an exciton, created by the window,
enlarges, and the slope becomes more steep. In the region below the
window perimeter inside the well round potential well for excitons
arises (it is showed by the arrow in Fig.~2). The exciton energy in
this well is lower, than the energy in regions, remote from the
window. The depth of the well increases, if the plane of the quantum
well approaches to the window. The appearance of the well is related
with charge origination nearby edges of the window. In the work
\cite{GorbTim2} the shift in the emission spectrum of indirect
excitons from the regions below the window perimeter was observed,
and this is probably related with the formation of the
aforementioned round potential well.

\section{Exciton density equation}

We assume the time of post-excitation establishment of quasilocal
equilibrium of electrons and holes and binding into excitons to be
much smaller than the time of the exciton lifetime and the time of
equilibrium establishment between different regions. In this case,
the free energy of quasilocal state can be regarded as a function of
the exciton density. A phenomenological equation for the exciton
density $n$ can be written down as
\begin{equation}\label{ex-main}\frac{\partial n}{\partial t}=
-\mbox{div}{\bf j}+G-\frac{n}{\tau},\end{equation}

\N where $G\left({\bf r}\right)$ is the pumping (the number of
excitons, created in a unit area in a unit time), $\tau$ is the
exciton lifetime, ${\bf j}=-M\nabla\mu$ is the exciton current
density, where $\mu$ is the chemical potential, $M$ is the exciton
mobility. We shall use for $M$ the Einstein condition $M=nD/{kT}$.
The corrections for the condition caused by Bose statistics of
excitons were studied in \cite{Ivanov} for quantum wells. But for
the temperature and the exciton density under investigations these
corrections are not essential. The chemical potential can be
expressed by free energy: $\mu=\delta F/{\delta n}$. We chose the
free energy in the Landau model:
\begin{equation}\label{F}F[n]=\int{d{\bf r}\left[\frac{K}{2}
{(\nabla n)}^2+f(n)+nV\right]}.\end{equation}

\N The term $\frac{K}{2}(\nabla n)^2$ characterizes the energy of
the non-homogeneity. Exciton additional energy in a nonuniform
electric field is taken into account with the term $nV$. According
to Landau method, we expand the free energy in a power series of
$(n-n_{\rm c})$ in the vicinity of its minimum:
\begin{equation}\label{f}f(n)=f(n_{\rm c})+\frac{a}{2}\left( n-n_{\rm
c}\right)^2+\frac{b}{4}\left(n-n_{\rm c}\right)^4.\end{equation}

\N The parameters $a$, $b$ and $n_{\rm c}$ in (\ref{f}) are
phenomenological and may be obtained from quantum-mechanics
calculations of the free energy of the exciton system at infinite
exciton lifetime approximation or comparing the theory with the
experiment. After the substitution (\ref{F}) and (\ref{f}) in
Eq.~(\ref{ex-main}), the latter is reduced to
\begin{equation}\label{ex-1}\frac{\partial n}{\partial t}=
\frac{D}{kT}\nabla\left[n \nabla\left({\frac{df}{dn}-K\Delta n}
\right)+n\nabla V\right] +G-\frac{n}{\tau}.\end{equation}

\N We must remark that for small densities in the free energy
(\ref{F}) the term $kTn\left(\ln n-1\right)$ has to be added, which
leads in the Eq.~(\ref{ex-1}) to the well-known term $D\Delta n$.
But its contribution is inessential at great exciton densities.

Eq.~(\ref{ex-1}) is a non-linear 2D phenomenological equation, which
describes density distribution of interacting excitons of high
concentration taking into account the pumping and finite lifetime.
We shall solve it instead of Gross-Pitayevsky equation \cite{GrPit},
because the wave function loses coherence at distances of order or
smaller than the distance between excitons. The finite lifetime and
the presence of constant exciton generation lead to new qualitative
peculiarities in the process of phase formation in comparison with
phase formation for stable particles. We assume, that
Eq.~(\ref{ex-1}) is also applied, if a condensed phase is
electron-hole liquid. Then, $n$ in the free energy (\ref{F}) is the
density of electron-hole pairs.

For numeric calculations let the units of length, concentration and
time be
\begin{equation}\label{units}l_{\rm u}=\sqrt{\frac{K}{-a}},
\,\,\,\,\, n_{\rm u}=\sqrt{\frac{-a}{b}},\,\,\,\,\, t_{\rm
u}=\frac{kTK}{Da^2n_{\rm u}},\end{equation}

\N correspondingly. Then, the generation rate and the energy are
measured in the units of $G_{\rm u}=n_{\rm u}/t_{\rm u}$ and $V_{\rm
u}=-a n_{\rm u}$, correspondingly. In dimensionless variables the
exciton density equation (\ref{ex-1}) becomes
\begin{equation}\label{ex-math}\frac{\partial n}{\partial t}=
\nabla\left[n\nabla\Bigl(-\Delta n+(3n_{\rm c}^2-1)n-3n_{\rm
c}n^2+n^3\Bigr)\right. \nonumber\\ \Bigl. +n\nabla
V\Bigr]+G-\frac{n}{\tau}.\end{equation}

\section{Calculations and results discussion}

Eq.~(\ref{ex-math}) was solved numerically in a 2D system in the
shape of a rectangular plate, the sizes of which exceeds much the
window radius, so, at the further increasing of window sizes, the
quasiparticle density distribution does not almost change. The
following values of the parameters of the system were chosen:
$\tau=10$~ns, $T_{\rm c}=5.7$~K, $n_{\rm c}=1.2n_{\rm
u}=3.33\cdot10^{10}$~cm$^{-2}$, $K n_{\rm c}^2=15.7$~meV, $-a n_{\rm
c}=0.826$~meV, $b n_{\rm c}^3=1.70$~meV. The free path length is
$l=\sqrt{D \tau}=1.41$~$\mu$m. The pumping $G$ is constant on the
disc with the radius $r_0$ and equals to zero outside the disc. An
example of the stationary solution of Eq.~(\ref{ex-math}) for the
exciton density is shown in Fig.~3.

\epsfxsize=0.8\columnwidth\centerline{\epsffile{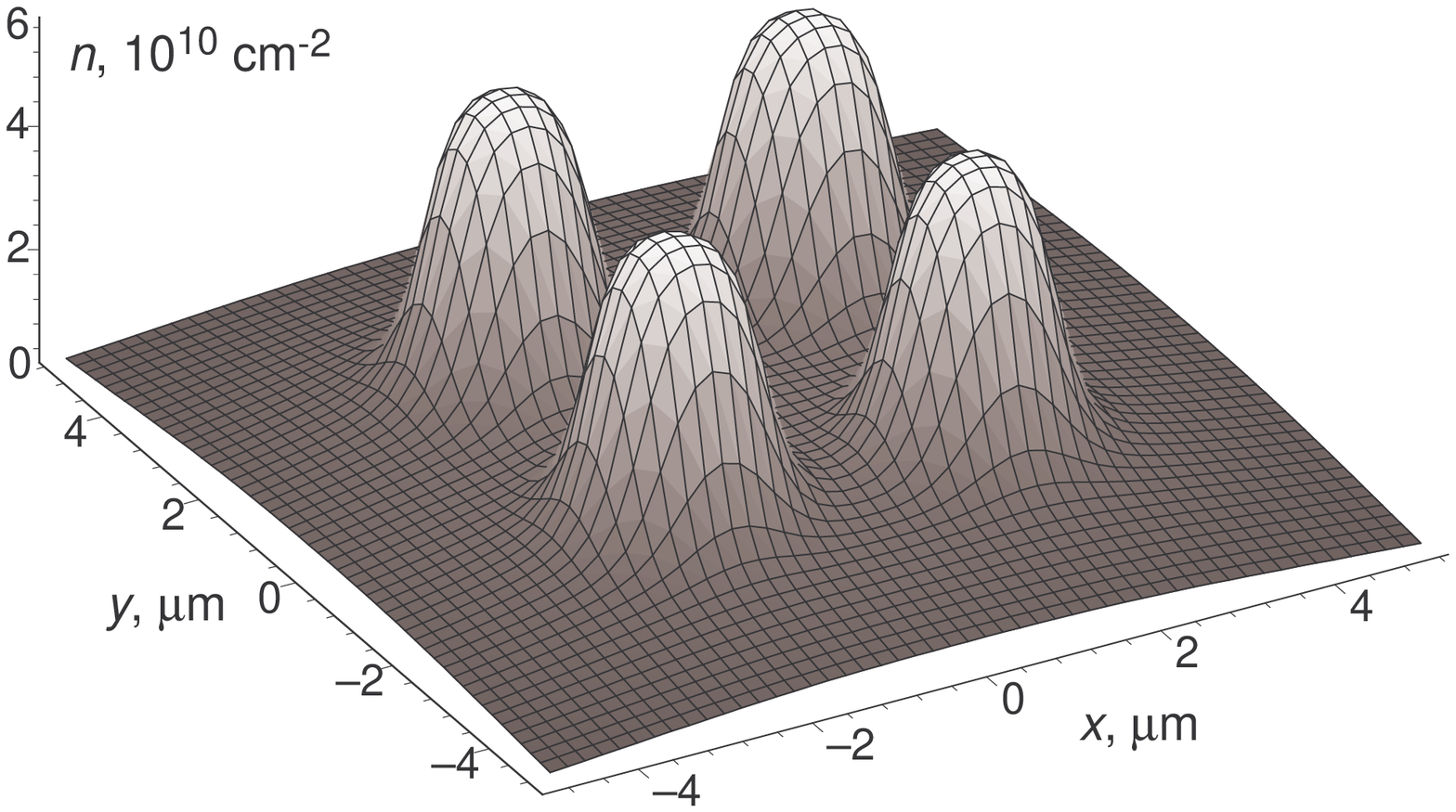}}

\N{\small Fig.~3. Exciton density distribution $n(x,y)$ in the
quantum well plane at the parameters: $r_0=2.5$~$\mu$m,
$z/r_0=1.44$, $T=1.71$~K, $G\tau=9.71\cdot10^{10}$~cm$^{-2}$,
$\lambda=30$}\vskip1mm

\N The points of high density can be attributed to a condensed
phase, and the points of low density are assumed to be a gas phase.
Thus, the formed fragments are periodically sited islands of the
exciton condensed phase, that corresponds to the structure, observed
in [16--18]. One can see from Fig.~3, that the islands of the
condensed phase appear at the boundary of the ring, as it is follows
from the qualitative analysis of the influence of nonuniform
potential distribution. A part of an island is under the metal mask
and may not be observed in the luminescence. The size of the region,
``hidden'' below the mask, depends on the distance of the quantum
well with respect to the upper electrode. With moving off the
electrode, the islands shift towards the window center and
``stretch'' from the electrode. With approaching the well plane to
the electrode islands ``hide'' below the electrode.

Hereinafter, we present the results of the investigation of the
structure behavior at external parameters changing. With broadening
the window, the number of islands increases (Fig.~4), as the length
of the circle, restricting the window, increases. At the same time,
the shape of the fragments does almost not change. At some small
window radius only the spot in the center forms, but the critical
value of the pumping, necessary for its appearance, is greater, than
the critical value for islands formation in the systems with bigger
window.

\epsfxsize=\textwidth\centerline{\epsffile{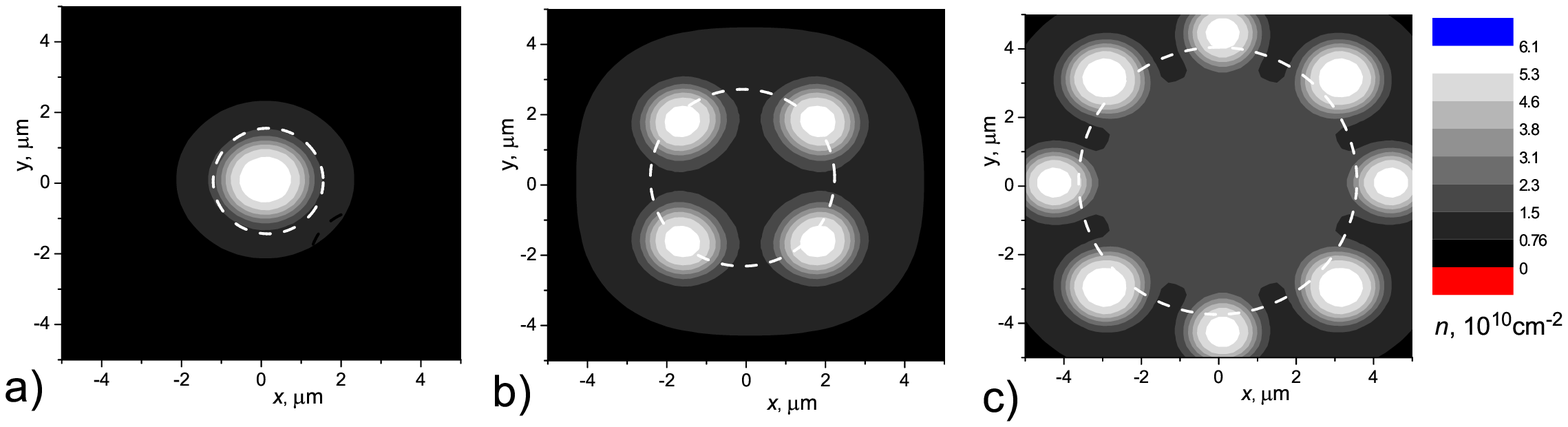}}

\N{\small Fig.~4. Exciton density distribution with growth of the
window radius. The radius equals to: a) $r_0=1.4$~$\mu$m, b)
$r_0=2.5$~$\mu$m, c) $r_0=3.6$~$\mu$m. Other parameters are the same
as in Fig.~3, except for the case `a', where the pumping $G$ is
larger (see the text). The case `b' is a density plot of 3D image in
Fig.~3}

Islands are formed only at the pumping exceeding some threshold,
because the condensed phase forms at large exciton density. At small
pumping the structure does not appear, and at large irradiation
intensity distinct islands merge into a continuous condensed phase
ring (Fig.~5).

\epsfxsize=\textwidth\centerline{\epsffile{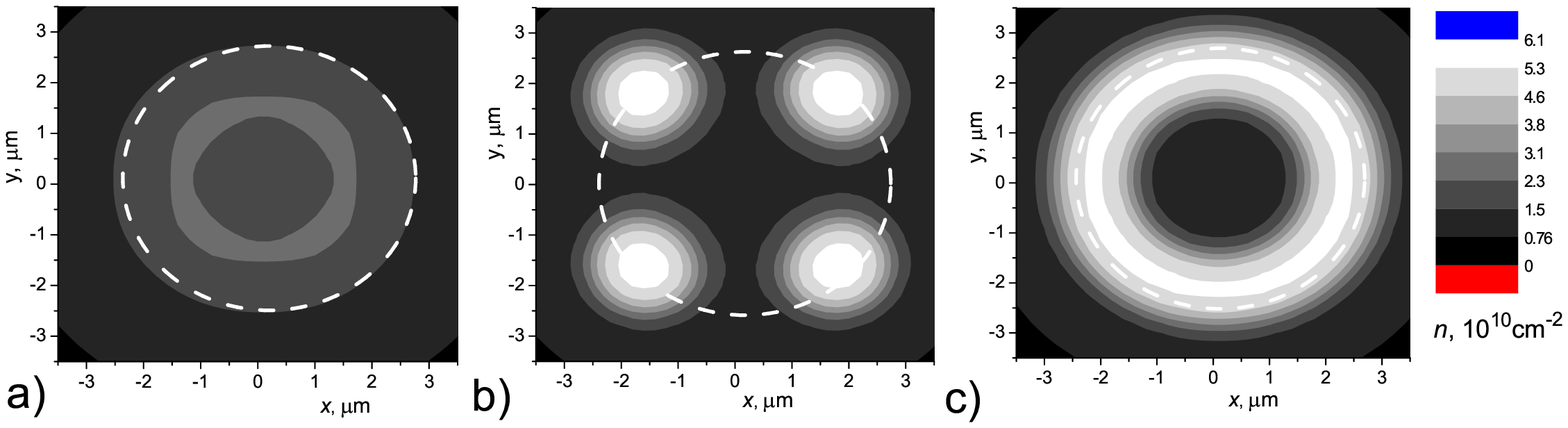}}

\N{\small Fig.~5. Exciton density distribution with the pumping
growth. The pumping $G$ equals to: a) $0.86G_0$, b) $G_0$, c)
$1.14G_0$; $G_0\tau=9.71\cdot10^{10}$~cm$^{-2}$. Other parameters
($r_0$, $z$, $T$) are the same as in Fig.~3}

Let us examine temperature dependence in the Landau model. The phase
transition occurs, if the coefficient $a$ in the free energy density
(\ref{f}) changes the sign. Temperature dependence of other
coefficients can be neglected. In Landau approximation, the
parameter $a$ depends on temperature like
\begin{equation}\label{a}a(T)=\alpha\left(1-\frac{T}{T_{\rm c}}
\right),\end{equation}

\N where $T_{\rm c}$ is the critical temperature, $\alpha<0$. Linear
dependence (\ref{a}) is real in the framework of self-consistent
field approximation and may not be fulfilled, if fluctuations play
the essential role. At derivation of the equation in dimensionless
variables, we shall use the units (\ref{units}), but substituting
$a$ and $T$ for $\alpha$ and $T_{\rm c}$, correspondingly. If
temperature is measures in $T_{\rm c}$, then in new variables the
exciton density equation (\ref{ex-math}) is modified to:
\[\label{ex-math2}\frac{\partial n}{\partial t}=
\frac{1}{T}\nabla\left[n\nabla\Bigl(-\Delta n+(3n_{\rm
c}^2-1+T)n-3n_{\rm c}n^2 \right. \Bigl. +n^3 \Bigr)
 +n\nabla V\Bigr]+G- \frac{n}{\tau}.\]

\N As numeric calculations show, with increasing the temperature
distinct islands of the condensed phase merge into a continuous
ring, and at higher temperatures the emission from the window is
homogeneous (Fig.~6).

\epsfxsize=\textwidth\centerline{\epsffile{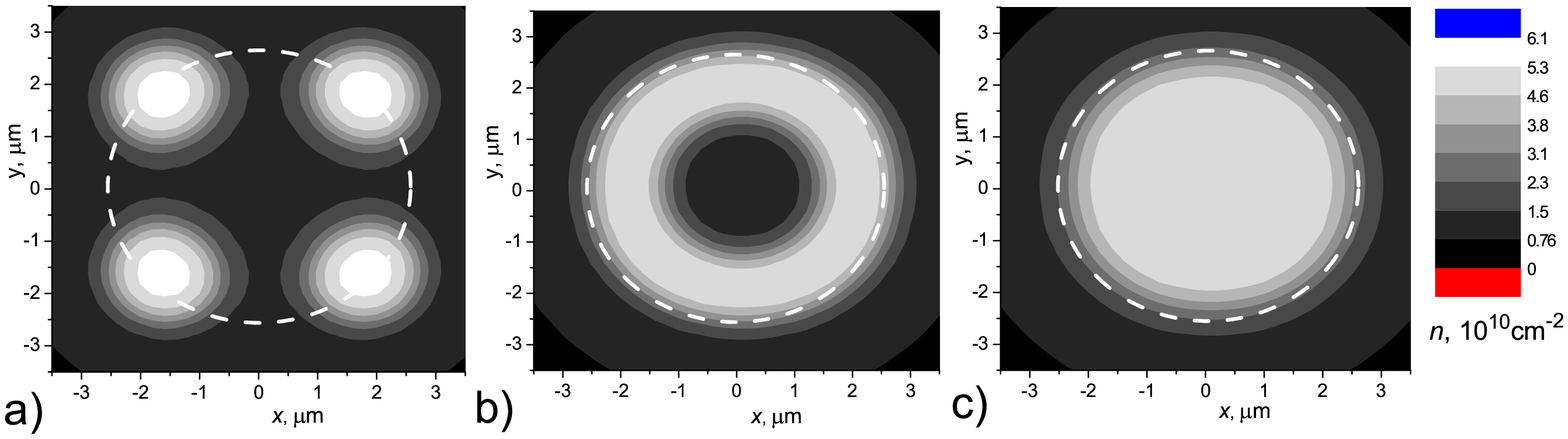}}

\N{\small Fig.~6. Exciton density distribution with temperature
increasing. Temperature equals to: a) 1.71~K, b) 1.94~K, c) 3.14~K.
Other parameters ($G$, $r_0$, $z$) are the same as in Fig.~3}

\section{Conclusions}

In the paper the investigation of exciton condensation in the
framework of traditional phase transition theory performed for the
system in nonuniform electric field taking into account the effects
of a finite value of the exciton lifetime and the presence of the
pumping. The main results are the following. 1) At the pumping,
exceeding some threshold value, the exciton condensed phase appears
in the form of the islands localized nearby the edge of the window
or continuous rings. 2) The number of the islands increases with the
rise of window radius. A continuous ring can break out into distinct
periodically sited condensed phase islands. 3) At small size of the
window, only the spot in the window center appears. 4) With the rise
of pumping and temperature the distinct islands merge into
continuous ring.

All above-mentioned peculiarities were observed in the works
[16-18]. We have to remark, that Bose-Einstein condensation was not
drawn for results obtained, but quantum statistics of excitons may
play some role in the formation of parameters, which were used in
the phenomenological model. We consider, that the formed structure
(the formation of islands, their location, dynamics at parameters
changing) is the consequence of non-equilibrium of the system, i.e.,
the structure is an example of self-organization processes in
non-equilibrium systems \cite{Haken,Prigozhin}. In the considered
case the origin of a non-equilibrium state is the finite value of
the exciton lifetime. The general theory of spatial structure
formation for unstable particles at phase transitions is presented
in \cite{SugSyn,SugSSC}.


\begin{thebibliography}{30}
\bibitem{MoscSnoke}S.A.~Moskalenko and D.W.~Snoke, {\it Bose-Einstein
condensation of excitons...} (Cambridge University Press, Cambridge,
2000).

\bibitem{LarTim}A.V.~Larionov and V.B.~Timofeev,
Pis'ma Zh. Eksp. Teor. Fiz. {\bf73}, 342 (2001).

\bibitem{Tim2}A.A.~Dremin, A.V.~Larionov, and V.B.~Timofeev, Fiz. Tverd.
Tela {\bf46}, 168 (2004). V.B.~Timofeev, Usp. Fiz. Nauk {\bf175},
315 (2005).

\bibitem{ButovNat}L.V.~Butov, A.C.~Gossard, and D.S.~Chemla,
Nature {\bf418}, 751 (2002); arXiv: cond-mat/0204482. L.V.~Butov,
Solid State Commun. {\bf127}, 89 (2003).

\bibitem{SnokeNat}D.~Snoke, S.~Denev, Y.~Liu et al., Nature {\bf418}, 754
(2002). D.~Snoke, Y.~Liu, S.~Denev et al., Solid State Commun.
{\bf127}, 187 (2003).

\bibitem{ButovPRL05}L.S.~Levitov, B.D.~Simons, and L.V.~Butov, Phys.
Rev. Lett. {\bf94}, 176404 (2005); arXiv: cond-mat/0503628.

\bibitem{Liu}C.S.~Liu, H.G.~Luo, and W.C.~Wu, J. Phys.:
Condens. Matter. {\bf18}, 9659 (2006); arXiv: cond-mat/0604564.

\bibitem{Parask}A.~Paraskevov and T.V.~Khabarova, arXiv: cond-mat/0611258.

\bibitem{Sug86}V.I.~Sugakov, Fiz. Tverd. Tela {\bf21}, 562 (1986).

\bibitem{Sug3}V.I.~Sugakov, Ukr. Fiz. Zhurn. {\bf49}, 1117 (2004); arXiv:
cond-mat/040739; Solid State Commun. {\bf134}, 63 (2005).

\bibitem{Sug4}V.I.~Sugakov, Fiz. Tverd. Tela {\bf48}, 1868 (2006);
Fiz. Nizk. Temper. {\bf32}, 1449 (2006).

\bibitem{Ch}A.A.~Chernyuk and V.I.~Sugakov, Phys. Rev. B
{\bf74}, 085303 (2006); arXiv: cond-mat/0508282.

\bibitem{ChP}A.A.~Chernyuk and V.I.~Sugakov, Acta Physica
Polonica A {\bf110}, No.2, 169 (2006).

\bibitem{LozBerm}Yu.E.~Lozovik and O.L.~Berman, Pis'ma Zh. Eksp.
Teor. Fiz. {\bf64}, 573 (1996); Zh. Eksp. Teor. Fiz. {\bf111}, 1879
(1997).

\bibitem{BermLozSnoke}O.L.~Berman, Yu.E.~Lozovik, and D.W.~Snoke
et al., Phys. Rev. B {\bf70}, 235310 (2004); arXiv:
cond-mat/0408053.

\bibitem{GorbTim}A.V.~Gorbunov and V.B.~Timofeev, Pis'ma Zh.
Eksp. Teor. Fiz. {\bf83}, 146 (2006).

\bibitem{GorbTim1}A.V.~Gorbunov and V.B.~Timofeev, Usp. Fiz.
Nauk {\bf176}, No.6, 652 (2006).

\bibitem{GorbTim2}A.V.~Gorbunov and V.B.~Timofeev, Pis'ma Zh.
Eksp. Teor. Fiz. {\bf84}, 390 (2006).

\bibitem{LL}L.D.~Landau and E.M.~Lifshits, {\it Theoretical
physics}, Vol.VIII, p.47 (Nauka, Moscow, 1982).

\bibitem{Ivanov}A.L. Ivanov, Europhys. Lett. {\bf
59}, 586 (2002); arXiv: cond-mat/0206459.

\bibitem{GrPit}E.N.~Gross, Nuovo Cimento {\bf20}, 454 (1961).
L.P.~Pitayevsky, Zh. Eksp. Teor. Fiz. {\bf13}, 451 (1961).

\bibitem{Haken}G.~Haken, {\it Synergetics} (Moskow, 1980).

\bibitem{Prigozhin}G.~Nicolis and I.~Prigozhin, {\it Self-organization
in non-equilibrium systems} (Wiley, New-York, 1977).

\bibitem{SugSyn}V.I.~Sugakov, {\it Lectures in synergetics} (World
Scientific, Singapore, 1998).

\bibitem{SugSSC}V.I. Sugakov, Solid State Commun. {\bf106}, 705 (1998).

\end{thebibliography}
\end{document}